\documentclass[twocolumn,showpacs,preprintnumbers,amsmath,amssymb]{revtex4}

\usepackage{graphicx}
\usepackage{dcolumn}
\usepackage{bm}

\begin{document}

\preprint{APS/123-QED}

\title{Detect overlapping and hierarchical community structure
in networks}
\author{Huawei Shen $^{1,2}$}
\author{Xueqi Cheng $^{1}$}\email{cxq@ict.ac.cn}
\author{Kai Cai $^{1}$}
\author{Mao-Bin Hu $^{3}$}
\affiliation{$^{1}$Institute of Computing Technology, Chinese
Academy of Sciences, Beijing, P.R. China \\
$^{2}$ Graduate University of Chinese Academy of Sciences, Beijing, P.R. China \\
$^{3}$ School of Engineering Science, University of Science and
Technology of China, Hefei 230026, P.R. China}

\date{\today}

\begin{abstract}
Clustering and community structure is crucial for many network
systems and the related dynamic processes. It has been shown that
communities are usually overlapping and hierarchical. However,
previous methods investigate these two properties of community
structure separately. This paper proposes an algorithm (EAGLE) to
detect both the overlapping and hierarchical properties of complex
community structure together. This algorithm deals with the set of
maximal cliques and adopts an agglomerative framework. The quality
function of modularity is extended to evaluate the goodness of a
cover. The examples of application to real world networks give
excellent results.

\end{abstract}

\pacs{89.75.Hc, 05.10.-a, 87.23.Ge, 89.20.Hh}

\maketitle

\section{Introduction}
\label{introduction}

Many complex systems in nature and society can be described in terms
of networks or graphs. Examples include the Internet, the
world-wide-web, social and biological systems of various kinds, and
many others \cite{Strogatz01, AB02, Newman03a}. In the past decade,
the theory of complex network has attracted much attention. Complex
networks are usually characterized by several distinctive
properties: power law degree distribution, short path length,
clustering and community structure. The problem becomes important
because complex system's dynamics is actually determined by the
interaction of many components and the topological properties of the
network will affect the dynamics in a very fundamental way.
Therefore, an efficient and sound approach that can capture the
topological properties of network is needed.

Identifying the community structure is crucial to understand the
structural and functional properties of the networks \cite{GN02,
GA05, FLGC02}. Many methods have been proposed to identify the
community structure of complex networks \cite{Newman06c, PDFV05,
NG04, Radicchi04, Newman04d, Newman06e}. One can refer to
\cite{DDDA05} for reviews. These methods can be roughly classified
into two categories in terms of their results, i.e., to form a
partition or a cover of the network.

The first kind of methods produce a partition, i.e each vertex
belongs to one and only one community and is regarded as equally
important. Different from classical graph-partition problem, the
number of communities and the size of each community are previously
unknown. Newman et al. proposed a quality function $Q$,
namely~\textit{modularity}, to evaluate the goodness of a partition
~\cite{NG04}. A high value of $Q$ indicates a significant community
structure. Several community detection methods have been proposed by
optimizing modularity~\cite{Newman04d, CNM04, DA05}. Generally, this
kind of methods are suitable to understand the entire structure of
networks, especially for the networks with a small size. Recently,
some authors~\cite{FB07, KSKK07} have pointed out that the
optimization of modularity has a fundamental drawback, i.e. the
existence of a resolution limit.

The second kind of methods aim to discover the vertex sets (i.e.
communities) with a high density of edges. In this case, overlapping
is allowed, that is, some vertices may belong to more than one
community. Meanwhile, some vertices may be neglected as subordinate
vertices. Therefore, these methods result in an incomplete cover of
the network. Numerous methods have been proposed, based on
\emph{k-clique}~\cite{PDFV05},~\emph{k-dense}~\cite{SYK06} or other
patterns. Unfortunately, there is no commonly accepted standard to
evaluate the goodness of a cover up to now. Compared to the
partition methods, this kind of methods are appropriate to find the
cohesive regions in large-scale networks.

In real networks, communities are usually overlapping and
hierarchical \cite{PDFV05,SGMA07,RSMOB02,P08}. Overlapping means
that some vertices may belong to more than one community.
Hierarchical means that communities may be further divided into
sub-communities. The two kinds of existing methods, as mentioned
above, investigate these two phenomena separately. The first kind of
methods can be used to explore the hierarchical community structure,
however, they are unable to deal with overlaps between communities.
The second kind of methods can uncover overlapping community
structure of networks, but they are incapable of finding the
hierarchy of communities. Recently, several authors begin to detect
the hierarchical and overlapping community structure~\cite{LFK08}.

In this paper, a new algorithm EAGLE (agglomerativE hierarchicAl
clusterinG based on maximaL cliquE) is presented to uncover both
hierarchical and overlapping community structure of networks. This
algorithm deals with the set of maximal cliques and adopts an
agglomerative framework. The effectiveness is demonstrated by
applications to two real-world networks, namely the word association
network and the scientific collaboration network.

\begin{figure*}
\begin{center}
\resizebox{16.8cm}{!}{\includegraphics{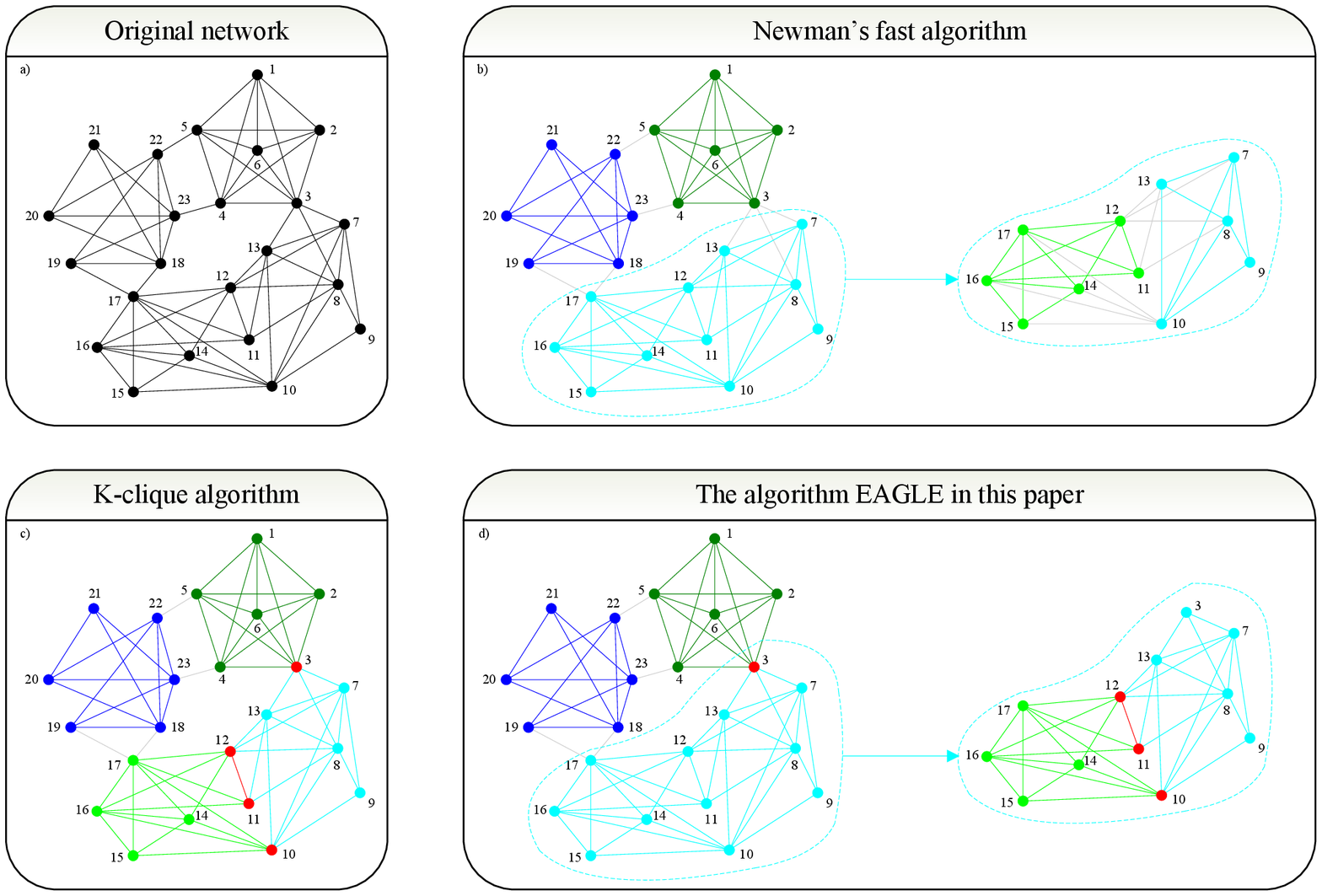}}
\end{center}

\caption{Comparison of community structure found by different
algorithms. Different communities are rendered in different colors.
Edges between communities are colored in light gray. Overlapping
region between communities are emphasized in red. a) The schematic
network. b) The hierarchical community structure found by Newman's
fast algorithm. This algorithm is chosen as a representative of the
first kind of algorithms. c) The overlapping community structure
found by the k-clique algorithm as a representative of the second
kind of algorithms. d) The hierarchical and overlapping community
structure found by the algorithm EAGLE.} \label{schematic}
\end{figure*}

In Fig.\ref{schematic}, we use a schematic network to illustrate
what EAGLE can do and compare it with the two kinds of existing
methods. Fig.\ref{schematic}(a) depicts the schematic network. We
construct this network according to the schematic network
in~\cite{PDFV05}, which has overlapped community structure. To
construct the hierarchy of the overlapped communities, we remove the
edge connecting vertices $9$ and $13$ and add two edges, one
connecting $10$ and $15$ and the other one connecting $10$ and $13$.
Fig.\ref{schematic}(b) shows the community structure found by
Newman's fast algorithm \cite{Newman04d}. Three community are found
when applying the algorithm to the schematic network. The hierarchy
of communities can be revealed by applying the algorithm to each
community further. For example, one of the three communities is
divided into two sub-communities. Overlaps between communities are
not allowed. Fig.\ref{schematic}(c) demonstrates the overlapping
community structure found by k-clique algorithm \cite{PDFV05}.
Unfortunately, this algorithm can not reveal the hierarchy of
community. Fig.\ref{schematic}(d) shows the hierarchical and
overlapping community structure found by our algorithm. EAGLE
provides a possible way to investigate a more complete picture of
the community structure.


\section{The algorithm: EAGLE}
\label{The algorithm}

A community can be regarded as a vertex set within which the
vertices are more likely connected to each other than to the rest of
the network. This indicates that a community usually has relatively
high link-density. Generally, the link-density of a clique is
highest among all kinds of vertex subsets of a network. Dense-linked
community usually contains a large clique, which could be regarded
as the core of the community. Based on this observation, the
algorithm EAGLE is proposed as an agglomerative hierarchical
clustering algorithm to investigate the community structure.
Different from traditional agglomerative algorithms
\cite{Newman04d}, our algorithm deals with the set of maximal
cliques rather than the set of sole vertices.

A \textit{maximal clique} is a clique which is not a subset of any
other cliques. In the algorithm EAGLE, we need to firstly find out
all the maximal cliques in the network. This can be done by many
efficient parallel algorithms. Here we choose the well-known
\emph{Bron-Kerbosch} algorithm \cite{BK73} for its simplicity in
implementation. Note that not all maximal cliques are taken into
account. The maximal cliques, whose vertices are from some other
larger maximal cliques, are called \textit{subordinate maximal
cliques}. For example, in Fig.\ref{schematic}, vertex $4$ and $23$
forms a subordinate maximal clique. Because vertex $4$ is from
another larger maximal clique \{1, 2, 3, 4, 5, 6\} and vertex $23$
is also from other larger maximal cliques, including \{18, 20, 21,
23\}, \{18, 20, 22, 23\} and \{18, 19, 22, 23\}. Subordinate maximal
cliques may mislead our algorithm and should be discarded. Most
subordinate maximal cliques have small sizes. Thus, we can discard
them by setting a threshold $k$ and neglecting all the maximal
cliques with the size smaller than $k$. This simple tactic may also
discard some non-subordinate maximal cliques. The higher the value
of $k$ is, the more non-subordinate maximal cliques are discarded by
mistake. On the other hand, the smaller the value of $k$ is, the
more subordinate maximal cliques are remained. In real world
networks, the threshold $k$ typically takes value between $3$ and
$6$. As to the network in Fig.\ref{schematic}, both $3$ and $4$ are
appropriate threshold values. As to the networks used in
Sec.\ref{applications}, $4$ is demonstrated to be an appropriate
threshold \cite{PDFV05}. After neglecting the maximal clique with
the size smaller than the threshold $k$, some vertices do not belong
to any remaining maximal cliques. We call these vertices as
\textit{subordinate vertices}.

Our algorithm have two stages. In the first stage, a dendrogram is
generated. In the second stage, we choose an appropriate cut which
breaks the dendrogram into communities. The first stage of the
algorithm EAGLE can be described as follows:
\begin{enumerate}
\item Find out all maximal cliques in the network. Neglect subordinate maximal
cliques. The remainders are taken as the initial communities. Each
subordinate vertex is also taken as an initial community comprising
the sole vertex. Calculate the similarity between each pair of
communities.

\item Select the pair of communities with the maximum
similarity, incorporate them into a new one and calculate the
similarity between the new community and other communities.

\item Repeat step $2$ until only one community remains.
\end{enumerate}

In the algorithm, the similarity $M$ between two communities $C_{1}$
and $C_{2}$ is defined as

\begin{equation}
M=\frac{1}{2m}\sum_{v\in C_{1},w\in C_{2},v\neq w}\biggl[
A_{vw}-\frac{k_{v}k_{w}}{2m}\biggr] . \label{Similarity}
\end{equation}

Here, $A_{vw}$ is the element of adjacency matrix of the network (We
only consider undirected, unweighted networks in this paper). It
takes value $1$ if there is an edge between vertex $v$ and vertex
$w$ and $0$ otherwise. $m=\frac{1}{2}\sum_{vw}{A_{vw}}$ is the total
number of edges in the network. $k_{v}$ is the degree of $v$.

\begin{figure}
\begin{center}
\resizebox{8cm}{!}{\includegraphics{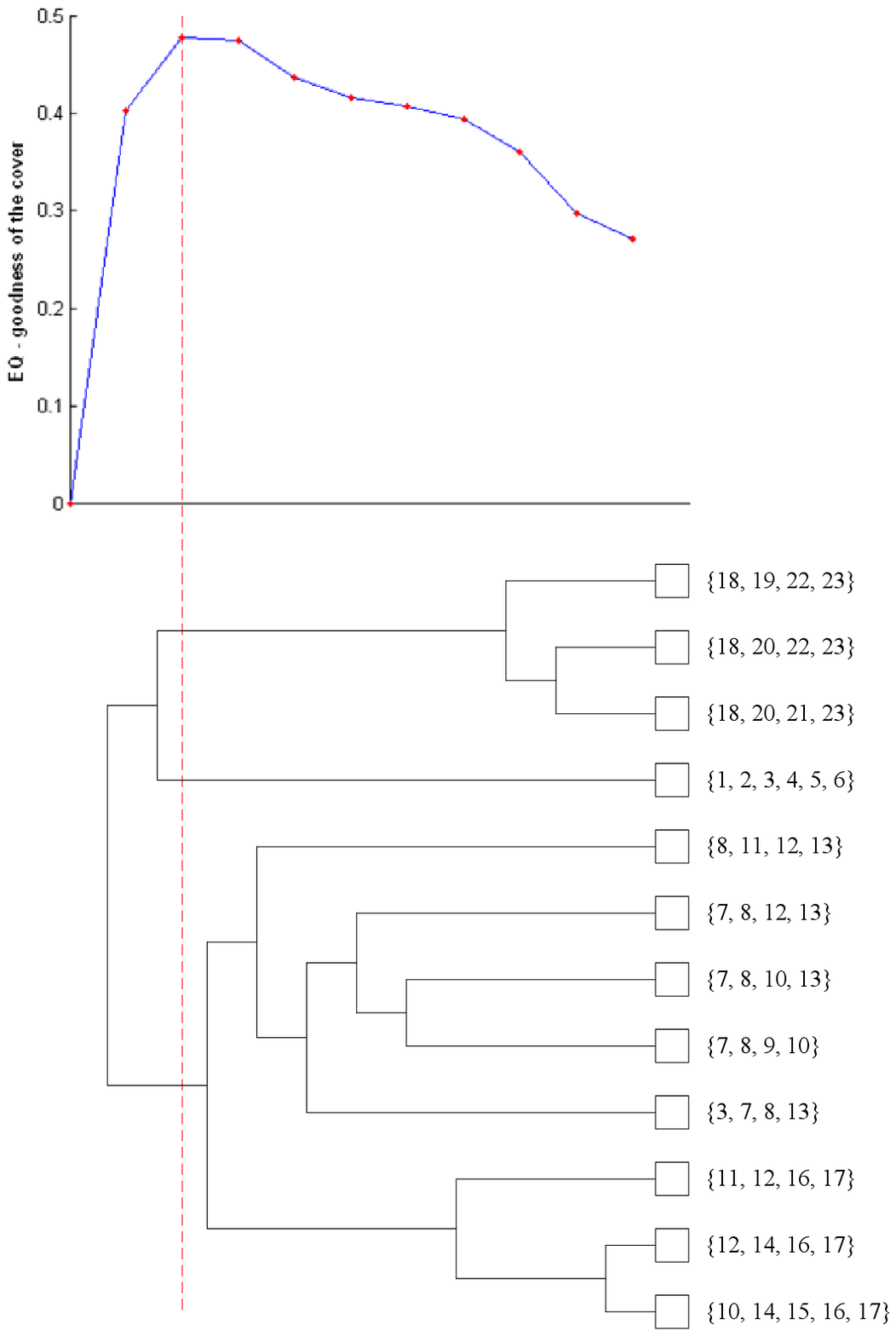}}
\end{center}
\caption{Illustration of the process of EAGLE when applied to the
schematic network in Fig.\ref{schematic}. The bottom part is a
dendrogram. The leaf nodes correspond to the non-subordinate maximal
cliques. The label of each leaf node shows the vertices belonging to
it. The red vertical dashed line is a cut through the dendrogram and
it gives the best cover of the network. The top part of the figure
is a graph which illustrates the curve of~$EQ$ corresponding to each
cover of the network. The threshold~$k$ is set to be~$4$.}
\label{SchematicDendrogram}
\end{figure}

Similar to the fast algorithm in \cite{Newman04d}, the process of
our algorithm corresponds to a dendrogram, which shows the order of
the amalgamations.Any cut through the dendrogram produces a cover of
the network. As an illustration, Fig.\ref{SchematicDendrogram} shows
the dendrogram generated by our algorithm when applied to the
network in Fig.\ref{schematic}.

The task of the second stage of the algorithm EAGLE is to cut the
dendrogram. To determine the place of the cut, a measurement is
required to judge the quality of a cover. In~\cite{NMCM08}, an
extension of modularity is proposed to evaluate the goodness of
overlapped community decomposition. In this paper, we propose
another extension of modularity $EQ$. As shown in
Fig.\ref{SchematicDendrogram}, the cut gives the best cover with the
maximum value of $EQ$. Given a cover of the network, let $O_{v}$ be
the number of communities to which vertex $v$ belongs. The extended
modularity is defined as
\begin{equation}
EQ=\frac{1}{2m}\sum_{i}\sum_{v\in C_i, w\in
C_i}\frac{1}{O_{v}O_{w}}\biggl[ A_{vw}-\frac{k_{v}k_{w}}{2m}\biggr]
. \label{eq}
\end{equation}
Note that $EQ$ reduces to $Q$ in \cite{NG04} when each vertex
belongs to only one community (Readers can refer to \cite{CNM04} for
details), and $EQ$ is equal to $0$ when all nodes belong to the same
community. In addition, it will be shown later in
Sec.\ref{applications}, a high value of $EQ$ indicates a significant
overlapping community structure.

Alike to modularity, the extended modularity suffers a resolution
limit beyond which no modular structure can be detected even though
these modules might have their own entity. As to EAGLE, however,
these modules can be still detected by further applying the
algorithm to each community found until none of them can be divided
into smaller ones. Thus, we obtain a hierarchy of overlapping
communities which reveals the community structure of network more
completely.

Then we analyze the time complexity of the algorithm. Let $n$ be the
number of vertices, $s$ be the number of maximal cliques in the
initial state of the algorithm, and $h$ be the number of pair of
maximal cliques which are neighbors (connected by edges or overlap
with each other). We firstly consider the first stage of the
algorithm. In step 1, $O(n^2)$ operations are needed to calculate
the similarity between each pair of initial communities. In step
$2$, we only consider the pairs of communities which are neighbors.
Each selection costs $h$ operations and each time of join costs
$O(n)$ operations at most. Totally, we carry on a maximum of $s-1$
join operations. Thus the first stage of the algorithm takes at most
$O(n^2+(h+n)s)$ operations. As to the second stage, we need to
calculate the value of $EQ$ corresponding to each cover. In our
implementation, we calculate the value of $EQ$ for the initial cover
and update it after each join of two selected communities into a new
one. Each time of update costs at most $n^2$ operations. Hence, the
second state of the algorithm takes at most $O(n^2s)$ operations. In
addition, we need to find out all the maximal cliques in the
network. It is widely believed to be a non-polynomial problem.
However, for real-world networks, finding all the maximal cliques is
easy due to the spareness of these networks.

Compared to the Newman's fast algorithm and the k-clique
algorithm, the algorithm EAGLE is time-consuming. We leave it as a
future work that how to improve the speed of EAGLE.

\section{Applications}
\label{applications}

In this section, we apply the algorithm EAGLE to two real-world
complex networks, the word association network and the scientific
collaboration network. The results show that EAGLE can discover new
knowledge and insights underlying these networks.

\begin{figure*}
\begin{center}
\resizebox{16.8cm}{!}{\includegraphics{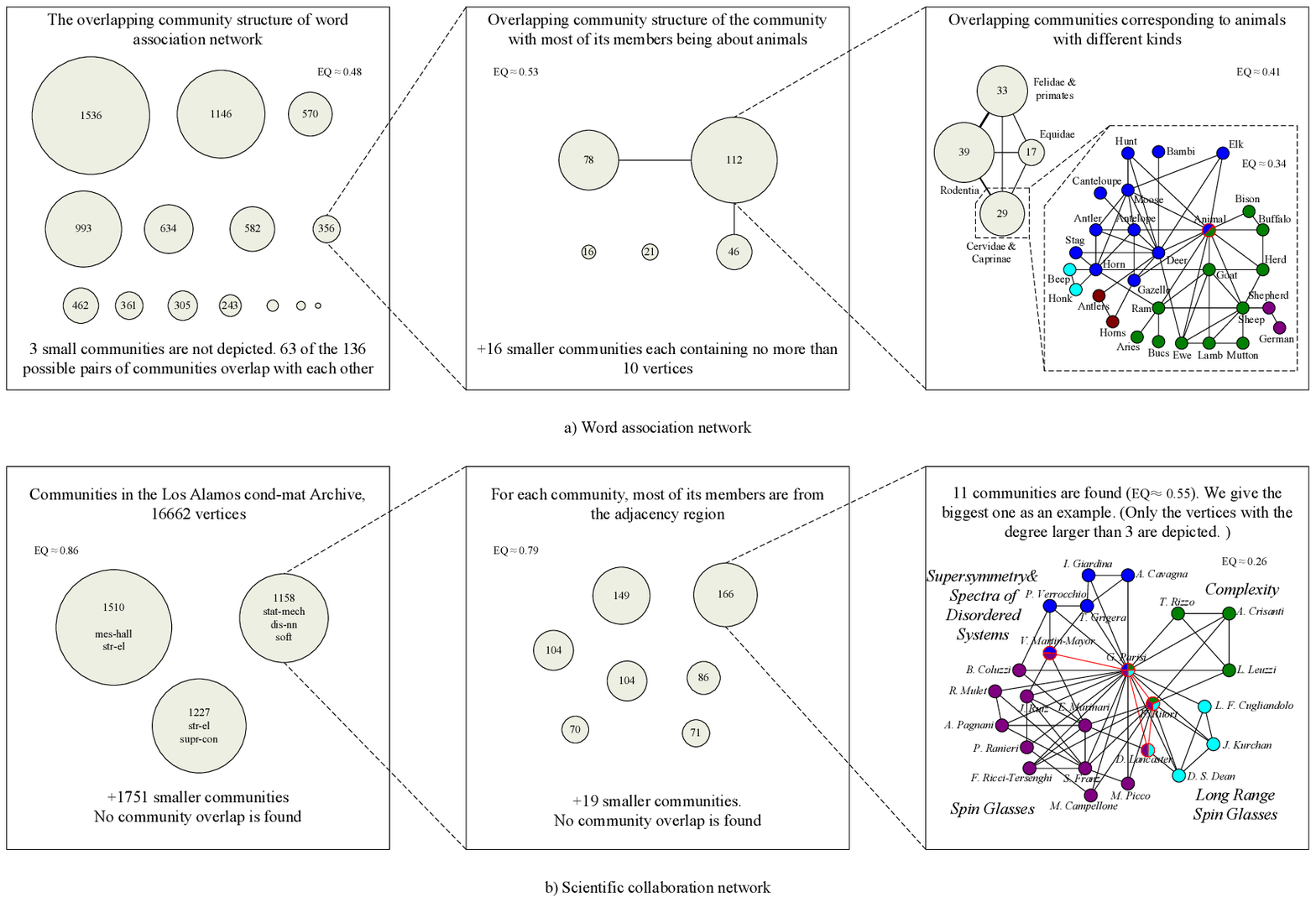}}
\end{center}
\caption{The hierarchical and overlapping community structure in a)
the word association network, and b) the scientific collaboration
network. Each numbered circle denotes a community and the number in
the circle denotes its size. Communities connected by a link overlap
with each other. Different communities are rendered in different
colors. The overlapping nodes and edges between communities are
colored in red. In addition, the values of the corresponding $EQ$
are also given when breaking networks (communities) down into
communities (sub-communities).} \label{hierarchy}
\end{figure*}

The test data of the two networks are from the demo of the
\emph{CFinder} \footnote{CFinder is a free software for finding
overlapping dense groups of nodes in networks, based on the Clique
Percolation Method (CPM) \cite{PDFV05}.}. The two networks comprise
$7207, 16662$ nodes and $31784, 22446$ edges, respectively. The
average clustering coefficients \cite{WS98} are approximately $0.15$
and $0.19$, which indicate that these networks have significant
community structures in general.

\begin{figure*}
\begin{center}
\resizebox{16.8cm}{!}{\includegraphics{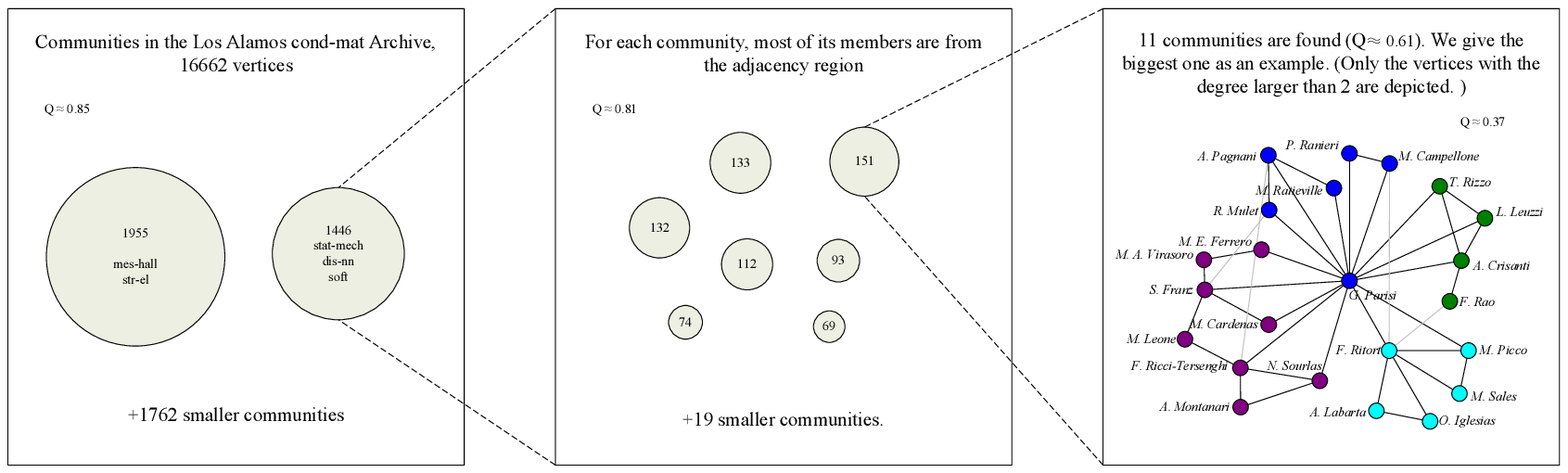}}
\end{center}
\caption{The hierarchical community structure found by Newman's fast
algorithm in the scientific collaboration network. Each numbered
circle denotes a community and the number in the circle denotes its
size. Communities connected by a link overlap with each other.
Different communities are rendered in different colors. The
overlapping nodes and edges between communities are colored in red.
In addition, the values of the corresponding $Q$ are also given when
breaking networks (communities) down into communities
(sub-communities).} \label{Newman-hierarchy}
\end{figure*}

\begin{figure}
\begin{center}
\resizebox{8cm}{!}{\includegraphics{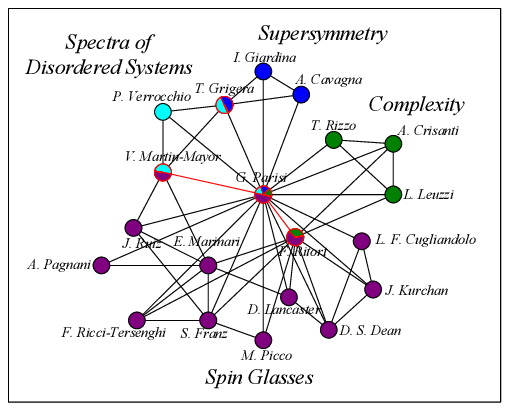}}
\end{center}
\caption{The overlapping community structure around the node G.
Parisi in the scientific collaboration network. Different
communities are rendered in different colors. The Overlapping nodes
and edges between communities are colored in red. Here, k is set to
be 4.} \label{clique_percolation}
\end{figure}

The word association network is constructed from the South Florida
Free Association norms list. The original network is directed and
weighted. The weight of a directed link from one word to another
indicates the frequency that the people in the survey associated the
end point of the link with its start point. The directed links are
replaced by undirected ones with a weight equal to the sum of the
weights of the corresponding two oppositely directed links.
Furthermore, the links with weight less than $0.025$ are deleted.
The scientific collaboration network is from the co-authorship
network of \emph{Los Alamos e-print} archives. Each article in the
archive between April 1998 and February 2004 contributes the value
$1/(n-1)$ to the weight of the link between every pair of its $n$
authors. The link with weight less than $1.0$ is omitted.

In the word association network, totally $17$ communities are found
by our algorithm - see Fig.\ref{hierarchy}(a), left panel. Among
these communities, $63$ of $136$ possible pairs of communities
overlap with each other. To investigate what is correlated to the
community structure, we apply our algorithm to each of these
communities again. The sub-community structure of one community is
given in Fig.\ref{hierarchy}(a), middle panel. Each of these
sub-communities have certain correlation with the semantic meaning
of words. For example, most of the words in the community with size
$112$ are related to the family of animals in Africa. This community
is explored further and four communities are found, shown in
Fig.\ref{hierarchy}(a), right panel. Each community is associated
with animals from the same family, namely rodentia, felidae \&
primates, cervidae \& caprinae, and equidae respectively. The
details of one community are also illustrated in
Fig.\ref{hierarchy}(a), right panel. Two large communities
correspond to words associated with animals from cervidae and
caprinae respectively. The overlapped word \textit{Animal} acts as a
bridge between the two communities. Three small communities comprise
peripheral words.

Applying our algorithm to the scientific collaboration network, we
obtain totally $1754$ communities - see Fig.\ref{hierarchy}(b), left
panel, with the corresponding high value of $EQ \approx 0.85$. Three
large communities contains $23.4\%$ of all the vertices, while the
others are relatively small. The three large communities correspond
closely to subject subareas: the biggest one mainly to
\textit{mes-hall} and \textit{str-el}, the second biggest one to
\textit{str-el} and \textit{supr-con}, and the other to
\textit{stat-mech}, \textit{dis-nn} and \textit{soft}. We further
apply the algorithm to one community and it is broken down into $26$
sub-communities - depicted in Fig.\ref{hierarchy}(b), middle panel.
There appears to be a correlation between the sub-community
structure and the regional divisions of the scientific researchers.
For example, most of the members of the community with size $166$
work in Europe. More specific regional information can be obtained
when applying the algorithm to this community. The biggest one and
its sub-community structure are given in Fig.\ref{hierarchy}(b),
right panel. We can see that the author G. Parisi (who is well known
for having made significant contributions in different fields of
physics) acts as a hub in the community. Different communities can
be associated with his different fields of interest.

Now, we compare the algorithm EAGLE with Newman's fast algorithm and
the k-clique algorithm by applying them to the scientific
collaboration network. Figure \ref{Newman-hierarchy} shows that the
hierarchical community structure found by Newman's fast algorithm.
The number of communities at each level of the hierarchy is almost
identical to that found by the algorithm EAGLE although the size of
each community is somewhat different. Compare the left panel of
Fig.\ref{Newman-hierarchy} with that of Fig.\ref{hierarchy}(b), one
community disappears. Actually, it is divided into several other
smaller communities, which are not depicted. As to the right panels,
the details of communities were given. The node G. Parisi, acting as
a hub in Fig.\ref{hierarchy}, only appear in one community in
Fig.\ref{Newman-hierarchy}. The reason is that Newman's algorithm
gives rise to partitions of network, while the algorithm EAGLE
allows overlaps between communities. Note that overlap between
communities is a very common phenomenon in real networks and may
contribute to the evolvement of communities and the dynamics of
networks.

Figure \ref{clique_percolation} shows the overlapping community
structure around the node G. Parisi in the scientific
collaboration network. Compare to Fig.\ref{hierarchy}, both the
algorithm EAGLE and the k-clique algorithm can find the
overlapping community structure, although the overlapped
communities found by the two algorithm are somewhat different.
However, the algorithm EAGLE can give the hierarchy of these
overlapped communities compared to the k-clique algorithm. The
hierarchy of communities is useful to understand the community
structure of real world networks.

\section{Conclusions and discussions}
\label{conclusions}

In this paper, we propose an algorithm, namely EAGLE, to uncover
both the hierarchical and overlapping properties of community
structure in complex networks. This algorithm deals with the set of
maximal cliques and adopts an agglomerative framework. The
effectiveness of this algorithm is demonstrated by applications to
two real-world networks, namely the word association network and the
scientific collaboration network. Results also show that the
algorithm EAGLE provides a possible way to gain a more complete
picture of the community structure of networks. Note that only
un-weighted and undirected networks are considered in this paper. In
our further work, EAGLE will be generalized to the weighted and/or
directed networks. How to improve the e¡Àciency of EAGLE will also
be considered.

Our method can help to analyze the community structure of some very
large networks. It can also shed some light on understanding the
topological and dynamical behavior of some large technological,
social and biological network systems.

\begin{acknowledgments}
This work was funded by the $973$ National Basic Research Program of
China under grant number $2004CB318109$ and National Natural Science
Foundation of China under grant number $60873245$. The authors
gratefully acknowledge S.~Fortunato for helpful suggestions. The
authors also thank Shi Zhou for enlightening discussions.

\end{acknowledgments}

\end{document}